\documentclass[aps,prl,twocolumn,showpacs,eqsecnum,nofootinbib]{revtex4}

\usepackage{epsfig}
\usepackage{graphicx}
\usepackage{dcolumn}
\usepackage{amsmath}
\usepackage{enumerate}

\begin{document}

\title{
Halo independent comparison of direct dark matter detection data
}

\author{\mbox{Paolo Gondolo$^{1,2}$}
and
\mbox{Graciela B.  Gelmini$^{3}$}
}

\affiliation{
 \mbox{$^1$ Department of Physics and Astronomy, University of Utah,
   115 S 1400 E Suite 201, Salt Lake City, UT 84112, USA }
 \mbox{$^2$ School of Physics, Korean Institute of Advanced Studies (KIAS), Seoul 130-722, South Korea }
\mbox{$^3$  Department of Physics and Astronomy, UCLA,
 475 Portola Plaza, Los Angeles, CA 90095, USA}
\\
{\tt paolo.gondolo@utah.edu},
{\tt gelmini@physics.ucla.edu}
}


\vspace{6mm}
\renewcommand{\thefootnote}{\arabic{footnote}}
\setcounter{footnote}{0}
\setcounter{section}{1}
\setcounter{equation}{0}
\renewcommand{\theequation}{\arabic{equation}}

\begin{abstract} \noindent
We extend the halo-independent method of Fox, Liu, and Weiner to include energy resolution and efficiency with arbitrary energy dependence, making it more suitable for experiments to use in presenting their results. Then we compare measurements and upper limits on the direct detection of  low mass ($\sim\!\!10$ GeV) weakly interacting massive particles with spin-independent interactions, including the upper limit on the annual modulation amplitude from the CDMS collaboration. We find that isospin-symmetric couplings are severely constrained both by XENON100 and CDMS bounds, and that isospin-violating couplings are still possible at the lowest energies, while the tension of the higher energy CoGeNT bins with the CDMS modulation constraint remains. We find the CRESST II signal is not compatible with the modulation signals of DAMA and CoGeNT.
\end{abstract}

\maketitle 

\vspace{-0.5cm}

\section{I. Introduction}

The nature of dark matter is one of the fundamental problems of
physics and cosmology.  Weakly interacting massive particles (WIMPs), 
i.e.\ particles with weakly interacting cross sections
and masses in the GeV--10~TeV range,
 are  among the best motivated candidates for dark matter. Of particular interest is a low mass region,
$\sim10$ GeV, suggested by data from three direct dark matter experiments: DAMA~\cite{dama2010}, CoGeNT~\cite{Aalseth:2010vx, Aalseth:2011wp} and CRESST-II~ \cite{Angloher:2011uu}. DAMA and CoGeNT report annual modulations with the expected characteristics of a WIMP signal~\cite{Drukier}. CRESST-II observes an excess of events above their known background, excess which may be interpreted as due to dark matter WIMPs. Stringent upper limits have been placed on dark matter WIMPs by other direct detection experiments. The most stringent limits in the region of $\sim10$~GeV WIMPs come from the XENON10~\cite{Angle:2011th}, XENON100~\cite{Aprile:2011hi, Aprile:2012nq}, SIMPLE~\cite{Felizardo:2011uw}, and CDMS experiments~\cite{Ahmed:2010wy}. All but one of these limits result from an upper bound on the total unmodulated event rate. The exception is a recent result by the CDMS collaboration~\cite{UCLA-Serfass}, which has searched for an annual modulation in their data and, not finding it, has placed a stringent upper limit  on its amplitude.

\section{The halo independent comparison method}

Here we compare the above measurements and upper limits in a halo-model independent fashion. We concentrate on light WIMPs with spin independent (SI) interactions, and extend the halo-independent method of Fox, Liu, and Weiner \cite{Fox:2010bz}, later extensively employed in~\cite{Frandsen:2011gi}, by including  energy resolution, efficiency, and form factors with arbitrary energy dependence. The use of constant efficiencies and form factors over bins constituted a limitation of earlier versions of the method. In our form, the method can be used by any experiment to present their own results in a way that would allow for an immediate comparison between experiments in a halo-independent manner.

 Fox, Liu, and Weiner \cite{Fox:2010bz} presented their method for differential and total rates.
 Statistical analyses usually use rates integrated over energy intervals, e.g. when computing maximum gap limits.  When integrating the differential rates over  energy bins, the energy dependence of efficiencies and form factors within each bin must be taken into account.   Fox, Liu, and Weiner~\cite{Fox:2010bz} took efficiencies and form factors constant over the bin.  For binned rates, they evaluated them at the central energy of each bin. For the total rate, they evaluated them at the energies that minimize or maximize the ratio of the rates to be compared, depending on whether one is considering a putative signal or a constraint. They also included the energy resolution of DAMA by smearing with a Gaussian distribution but the details of this procedure are not spelled out in their paper. Having to use constant efficiencies and form factors lead Frandsen et al.~\cite{Frandsen:2011gi} to assume that each bin is sufficiently small so that these quantities do not vary significantly within a single bin. This is  a particularly limiting restriction in analyzing the CRESST II data due to the onset of detector module thresholds. Our method overcomes all of these limitations.
 
The differential recoil rate per unit detector mass,  typically  in units of
counts/kg/day/keV, for the scattering of WIMPs of mass $m$ off 
nuclei of mass number $A$, atomic number $Z$, and mass $m_{A,Z}$ is
\begin{equation} 
  \frac{dR_{A,Z}}{dE}  = \frac{\sigma_{A,Z}(E)}{2 m \mu_{A,Z}^2}\, \rho\, \eta(v_{\rm {min}},t) ,
  \label{dRdE}
\end{equation}
where $E$ 
is the nucleus recoil energy,  $\rho$ is the local WIMP density, $\mu_{A,Z} = m~m_{A,Z}/ (m + m_{A,Z})$ is the WIMP-nucleus reduced mass,
 $\sigma_{A,Z}(E)$ is (a multiple of)
the WIMP-nucleus differential cross-section $d\sigma_{A,Z}/dE= \sigma_{A,Z}(E)~m_{A,Z}/  2\mu_{A,Z}^2 v^2$, and
\begin{align}
\eta(v_{\rm min},t) = \int_{|{\bf v}|>v_{\rm min}} \frac{f({\bf v},t)}{v} d^3 v
\end{align}
is a velocity integral carrying the only dependence on the (time-dependent) distribution $f({\bf v},t)$ of WIMP
velocities ${\bf v}$ relative to the detector. Here
\begin{equation}
 v_{\rm {min}} = \sqrt{\frac{m_{A,Z} E}{2\mu_{A,Z}^2}}
  \label{vmin}
\end{equation}
is the minimum WIMP speed that can result in a recoil energy $E$ in an elastic scattering with the $A,Z$ nucleus.
Due to the revolution of the Earth around the Sun, the $\eta$ function has an annual modulation
generally well approximated by the first  terms of a harmonic series
\begin{equation}
  \eta(v_{\rm {min}},t) = \eta_0(v_{\rm {min}}) + \eta_1(v_{\rm {min}}) \cos{\omega(t-t_0)},
   \label{eta} 
\end{equation}
where $\omega = 2\pi$/yr and $t_0$ is the time of maximum signal.

For spin-independent interactions (SI),  the WIMP-nucleus cross-section can be written in terms of the effective WIMP-neutron and WIMP-proton coupling constants $f_n$ and $f_p$ as
\begin{equation} 
  \sigma^{SI}_{A,Z}(E) =  \sigma_p \frac{ \mu_{A,Z}^2}{ \mu_p^2} [Z+ (A-Z)(f_n/f_p)]^2\,   \,F_{A,Z}^2(E) \, ,
  \label{sigma}
\end{equation}
where $\sigma_p$ is the WIMP-proton cross-section and 
$F_{A,Z}^2(E)$ is a nuclear form factor, which we take to be a Helm form factor~\cite{Helm:1956zz} normalized to $F_{A,Z}(0) = 1$.
In most  models the couplings are isospin conserving, $f_n = f_p$.
Isospin-violating couplings $f_n \ne f_p$ have been considered as a possibility to weaken the upper bounds  obtained with heavier target elements, which being richer in neutrons than lighter elements, have their couplings to WIMPs suppressed for $f_n/f_p \simeq -0.7~$\cite{Kurylov:2003ra}. 

Fox, Liu, and Weiner~\cite{Fox:2010bz} observed that the factor
\begin{equation} 
   \tilde\eta(v_{\rm min}) = \sigma_p (\rho/m) \eta(v_{\rm min})~,
     \label{tilde-eta}
\end{equation}
in Eq.~(\ref{dRdE}) with SI interactions is common to all experiments, and  compared direct detection experiments without any assumption about the dark halo of our galaxy by expressing the data in terms of $v_{\rm min}$ and $\tilde\eta(v_{\rm min})$. This was done extensively in~\cite{Frandsen:2011gi}, separately for $\tilde\eta_0(v_{\rm min})=\sigma_p (\rho/m) \eta_0$ and $\tilde\eta_1(v_{\rm min})=\sigma_p (\rho/m) \eta_1$. Since the $E$-$v_{\rm min}$ relation depends explicitly on the WIMP mass $m$, this procedure can be carried out only by fixing $m$ (except when $m$ is much smaller than the masses of all  nuclei involved, in which case the combination $m v_{\rm min}$ becomes independent of $m$). 

However most experiments do not measure the recoil energy $E$ directly, but rather a detected energy $E'$ subject to measurement uncertainties and fluctuations. These are expressed in an energy response function $G_{A,Z}(E,E')$ that incorporates the energy resolution $\sigma_E(E')$ and the mean value $\langle E'\rangle = E \, Q_{A,Z}(E)$, where $Q_{A,Z}(E)$ is the quenching factor. In this context, recoil energies are often quoted in keVnr, while detected energies are quoted in keVee (keV electron-equivalent) or  directly in photoelectrons. Moreover, experiments have an overall counting efficiency or cut acceptance $\epsilon(E')$ that depends on $E'$. A compound detector with mass fraction $C_{A,Z}$ in nuclide $A,Z$ has an expected event rate equal to
\begin{align}
\frac{dR}{dE'} = \epsilon(E') \, \int_0^\infty dE \, \sum_{A,Z} C_{A,Z} \, G_{A,Z}(E,E') \, \frac{dR_{A,Z}}{dE} .
\label{RSI-0}
\end{align}

We observe that the factor $\tilde\eta(v_{\rm min})$ is common to all experiments also when the rates are expressed in terms of the detected energies $E'$ as in Eq.~(\ref{RSI-0}). This observation allows us to extend Fox et al.'s method to the more realistic case of finite energy resolutions and $E'$-dependent efficiencies, without restrictions on how rapidly these quantities change with energy.

\section{II. Including energy resolution and efficiency with arbitrary energy dependence}

For this purpose, using $dE=(4\mu_{A,Z}^2/m_{A,Z}) v_{\rm min}  $ $dv_{\rm min}$, we write the average of Eq.~(\ref{RSI-0}) over a detected energy interval $[E'_1,E'_2]$ as
\begin{align}
R_{[E'_1,E'_2]} =  \int_0^\infty dv_{\rm min}  \,\, \mathcal{R}^{SI}_{[E'_1,E'_2]}(v_{\rm min})  \, \tilde\eta(v_{\rm min}) .
\label{RSI-2}
\end{align}
Here we have defined the response function for SI WIMP interactions, with $E_{A,Z}=2\mu_{A,Z}^2v_{\rm min}^2/m_{A,Z}$,
\begin{align}
\mathcal{R}^{SI}_{[E'_1,E'_2]}(v_{\rm min}) = &\sum_{A,Z}  \frac{2\,v_{\rm min}\,C_{A,Z} \, \sigma^{SI}_{A,Z}(E_{A,Z})}{m_{A,Z}\,\sigma_p\,(E'_2-E'_1)}  \times \nonumber \\&\quad  \int_{E'_1}^{E'_2} dE' \,   G_{A,Z}(E_{A,Z} ,E')   \, \epsilon(E') .
\label{Response}
\end{align}

When several energy bins are present, like when binning the data in energy or computing the maximum gap upper limit, we label each energy interval with an index $i$ and write
$R_{i}(t) $, 
$\mathcal{R}^{SI}_{i}(v_{\rm min})$, etc,\ for quantities belonging to the $i$-th energy interval. For example, binning the harmonic series in Eq.~(\ref{eta}) in energy gives
\begin{align}
R_i(t) = R_{0i} + R_{1i} \cos\!\left[ \omega (t-t_0) \right] .
\end{align}

Our task is to gain knowledge on the functions $\eta_0(v_{\rm {min}})$ and $\eta_1(v_{\rm {min}})$ from measurements $\hat R_{0i} \pm \Delta R_{0i}$ and $\hat R_{1i} \pm \Delta R_{1i}$ of $R_{0i}$ and $R_{1i}$, respectively. This is possible when a range of detected energies $[E'_1,E'_2]$ corresponds to only one range of $v_{\rm {min}}$ values  $[v_{\rm min,1},v_{\rm min,2}]$, for example when the measured rate is due to interactions with one nuclide only.   In this case, $[v_{\rm min,1},v_{\rm min,2}]$ is the $v_{\rm min}$ interval where  the response function $\mathcal{R}^{SI}_{[E'_1,E'_2]}(v_{\rm min})$ is significantly different from zero.  Ref.~\cite{Frandsen:2011gi} approximated this interval with $v_{\rm min,1} = v_{\rm min}(E'_1-\sigma_E(E'_1))$ and $v_{\rm min,2} = v_{\rm min}(E'_2+\sigma_E(E'_2))$.  When isotopes of the same element are present, like for Xe or Ge, the $v_{\rm min}$ intervals of the different isotopes almost completely overlap, and $v_{\rm min,1}$, $v_{\rm min,2}$ could be the $C_{A,Z}$-weighted averages over the isotopes of the element. When there are nuclides belonging to very different elements, like Ca and O in CRESST-II, a more complicated procedure should be followed (see below). 

Once the $[E'_1,E'_2]$ range has been mapped to a $[v_{\rm min,1},v_{\rm min,2}]$ range, we can estimate the $v_{\rm min}$-weighted averages 
\begin{align}
\overline{\tilde\eta_{[E'_1,E'_2]}} = \frac{\int_{v_{\rm min,1}}^{v_{\rm min,2}} \mathcal{R}^{SI}_{[E'_1,E'_2]}(v_{\rm min}) \, \tilde\eta(v_{\rm min}) \, dv_{\rm min}}{\int_{v_{\rm min,1}}^{v_{\rm min,2}} \mathcal{R}^{SI}_{[E'_1,E'_2]}(v_{\rm min})  \, dv_{\rm min}} 
\end{align}
as
\begin{align}
  \overline{\tilde\eta_{[E'_1,E'_2]}}  = \frac{\hat{R}_{[E'_1,E'_2]}}{\mathcal{A}^{SI}_{[E'_1,E'_2]}},
\end{align}
where
\begin{align}
\mathcal{A}^{SI}_{[E'_1,E'_2]} = \int_{v_{\rm min,1}}^{v_{\rm min,2}} \mathcal{R}^{SI}_{[E'_1,E'_2]}(v_{\rm min})  \, dv_{\rm min} .
\end{align}
In the case of binned data, these equations read $\overline{\tilde\eta_{0i}} = \hat{R}_{0i}/\mathcal{A}^{SI}_{i}$, $\overline{\tilde\eta_{1i}} = \hat{R}_{1i}/\mathcal{A}^{SI}_{i}$, with errors $\Delta\overline{\tilde\eta_{0i}} = \Delta{R}_{0i}/\mathcal{A}^{SI}_{i}$ and $\Delta\overline{\tilde\eta_{1i}} = \Delta{R}_{1i}/\mathcal{A}^{SI}_{i}$.

Upper limits on binned data can be set by replacing $\hat{R}_{[E'_1,E'_2]}$ above with the measured upper limit. Upper limits on unbinned data can be set using the method of Fox et al.~\cite{Fox:2010bz}, which we repeat here. The smallest non-increasing function $\eta(v_{\rm min})$ passing through a point $(v_s,\eta_s)$ is the downward step function $\eta(v_{\rm min})=\eta_s$ for $v_{\rm min}\le v_s$ and zero otherwise.  Using this  $\eta(v_{\rm min})$ function in Eq.~\ref{tilde-eta}, with  $\tilde\eta_s= \sigma_p (\rho/m)\eta_s$, the smallest event rate with $\eta(v_{\rm min}) = \eta_s$ at $v_{\rm min}=v_s$ is
\begin{align}
R^{\rm min}_{[E'_1,E'_2]} =   \tilde\eta_s \int_0^{v_s} dv_{\rm min}  \,\, \mathcal{R}^{SI}_{[E'_1,E'_2]}(v_{\rm min})  ,
\label{RSI-min}
\end{align}
where $[E'_1,E'_2]$ is any energy interval in which measurements of the rate have been done. This equation is used to bound the value of $\eta_s$ as a function of $v_s$.
We use it in the maximum gap method~\cite{Yellin:2002} for CDMS, XENON10, XENON100, and SIMPLE unbinned data.   When using the maximum gap method the choice of the $[E'_1,E'_2]$ interval is dictated by the data~\cite{Yellin:2002}.

For compound detectors like SIMPLE, when using Poisson or Binned Poisson statistics, Eq.~(\ref{RSI-min}) is equivalent but more transparent than the method in Appendix A.1 of~\cite{Frandsen:2011gi}. The main issue is that the number of events (or the upper bound on the number of events, depending on whether one is dealing with a measurement or a bound) is independent of the detector composition. In this case, the following relation holds between the value (or upper bound) $ \overline{\tilde\eta}$ for a compound and the value (or upper bound) $ \overline{\tilde\eta_{A,Z}}$ defined assuming only element $A,Z$ contributes to the rate:
\begin{align}
N =    \overline{\tilde\eta}  ~\mathcal{A}^{SI} =  \overline{\tilde\eta_{A,Z}} ~\mathcal{A}_{A,Z}^{SI}~,
\label{N}
\end{align}
where $N$ is a constant
Using that by definition  $\mathcal{A}^{SI} =  \sum_{A,Z}  \mathcal{A}_{A,Z}^{SI}$, and from Eq.~(\ref{N})  $\mathcal{A}_{A,Z}^{SI} = N/ ~ \overline{\tilde\eta_{A,Z}}$, we get, using Eq.~(\ref{N}) again, $N=   \overline{\tilde\eta} \sum_{A,Z} ( N/  ~\overline{\tilde\eta_{A,Z}}~)$ or
\begin{align}
\frac {1} {\overline{\tilde\eta}} =  \sum_{A,Z} ~\frac{1}{ \overline{\tilde\eta_{A,Z}}} ~,
\label{Eq-Frandsen}
\end{align}
which is Eq.(A.1) of~\cite{Frandsen:2011gi}.
This shows the equivalence of  the method in Appendix A.1 of~\cite{Frandsen:2011gi} with our simpler Eq.~(\ref{RSI-min}).

\section{III. Measurements of and limits on $ \tilde\eta_0$, $ \tilde\eta_1$}

The data and detector properties we use are as follows. (We acknowledge criticism of some experimental analyses~\cite{criticisms}, and try to be conservative.)

{\it CoGeNT.} We use the list of events,  quenching factor,  efficiency,  exposure times and  cosmogenic background  given in the 2011 CoGeNT data release~\cite{CoGeNT2011release}. We separate the modulated and unmodulated parts with a chi-square fit after binning in energy and in 30-day time intervals (we fix the modulation phase to DAMA's best fit value of 152.5 days from January 1). We correct the unmodulated part by surface-event correction factors $C(E)=1-e^{-E^2/E_C^2}$, which are similar to those in \cite{KelsoHooperBuckley} for $E_C=1.04$ keVee (``CoGeNT high"), 0.92 keVee (``CoGeNT med."), and 0.8 keVee (``CoGeNT low"). We leave it to the reader to subtract a possible constant background contribution $b_0$, since it is unknown. Thus, for CoGeNT we do not plot $\eta_0$ but $\eta_0 +b_0$ in the figures.

{\it CDMS.} For the upper limit on the total event rate we use only the T1Z5 detector~\cite{Ahmed:2010wy}, which gives the most stringent limits at low WIMP masses. The energy resolution is $[0.293^2+(0.056E)^2)]^{1/2}$, and the range for the maximum gap method is 2 keV--20 keV. For the modulation amplitude we use the 95\% upper bound of 0.045 events/kg-day-keV for a modulation phase equal to DAMA's in the energy range 5 keV--11.9 keV~\cite{UCLA-Serfass}.

{\it DAMA.} We read the modulation amplitudes from~\cite{dama2010}. We consider scattering off Na only, since the I component is under threshold for low mass WIMPs and reasonable local Galactic escape velocity. We show results for two values of the Na quenching factor: 0.3 and 0.45 (the latter suggested in \cite{Hooper:2010uy}). No channeling is included, as per~\cite{Bozorgnia:2010xy}.

{\it XENON100.} We show limits imposed by the last two data releases of Refs.~\cite{Aprile:2011hi} and  ~\cite{Aprile:2012nq} (with dashed and solid lines respectively). The exposure  in Ref.~\cite{Aprile:2011hi}  is 48 kg $\times$ 100.9 days. We convert the energies of the three candidate events in Ref.~\cite{Aprile:2011hi} into S1 values, and use the Poisson fluctuation formula Eq.~(15) in \cite{XENON100likelihood} to compute the energy fluctuations. We use the light efficiency function $\mathcal{L}_{\rm eff}$ in Fig.~1 of ~\cite{Aprile:2011hi}. We obtain the cut acceptance by multiplying two factors: the overall cut acceptance, which we set to a conservative value of 0.6 since it is unclear why in Fig.~2 of~\cite{Aprile:2011hi} it would depend on the WIMP mass when expressed as a function of S1, and the S1/S2 discrimination acceptance, taken from the just mentioned Fig. 2. We use a maximum gap method over the interval $4\le S1 \le 30$ photoelectrons.  In Ref.~\cite{Aprile:2012nq} the exposure is 34 kg $\times$ 224.6 days and there are two candidate events at lower energies than in the previous data set. Thus the new limits are less stringent at low energies and more stringent at higher energies than those derived from the previous data. For the new limits we use the same method just mentioned but with the  cut acceptance and data points of Ref.~\cite{Aprile:2012nq}. 

{\it XENON10.} We follow Ref.~\cite{Angle:2011th} and use only S2 without S1/S2 discrimination. The exposure is 1.2 kg $\times$12.5 days. We consider the 32 events within the 1.4 keV-10 keV acceptance box in the  Phys.\ Rev.\ Lett.\ article (not the arxiv preprint, which had an S2 window cut). We take a conservative acceptance of 0.94. For the energy resolution, we are more conservative than~\cite{Angle:2011th}: we convert the quoted energies into number of electrons $n_e=E \mathcal{Q}_y(E)$, with $\mathcal{Q}_y(E)$ as in Eq.~1 of~\cite{Angle:2011th} with $k=0.11$,  and use the Poisson fluctuation formula in~\cite{LewinSmith}.

{\it SIMPLE.} We consider only Stage 2, with an exposure of 6.71 kg days and no observed candidate event. We take an efficiency $\eta'(E)= 1 - \exp\{- \Gamma[(E/ E_{thr})-1]\}$   with $\Gamma= 4.2 \pm$0.3. With no events observed, the Poisson and maximum gap upper limits coincide.

{\it CRESST-II.} We take the histogram of events in Fig.~11 of Ref.~\cite{Angloher:2011uu}.  The acceptance is obtained by adding each module at its lower energy acceptance limit in their Table 1. The electromagnetic background is modeled as one $e/\gamma$ event in the first energy bin of each module. The exposure is 730 kg days. We assume a maximum WIMP velocity in the Galaxy such that W recoils can be neglected. To take into account the Ca and O components, we follow the same philosophy as Method 2 in Appendix A.2 of~\cite{Frandsen:2011gi}, but, without having to assume a constant efficiency in each energy bin, we are able to cover the CRESST-II energy range without gaps with the following binning: three high-energy bins ($i=4,5,6$) with scatterings off O only (assuming a maximum $v_{\rm min}$ of $\sim750$ km/s): $[17,20]$, $[20,23]$, and $[23,26]$ keV; and three corresponding low-energy bins ($i=1,2,3$) with the same $v_{\rm min}$ range and scatterings off O and Ca: $[11,13]$, $[13,15]$, and $[15,17]$ keV. To avoid complications with the overlap of the tails of the weight functions $\mathcal{R}^{SI}_{i}(v_{\rm min})$, we cut them outside the $v_{min}$ interval  $[v_{\rm min}(E'_{1i}),  v_{\rm min}(E'_{2i})]$, i.e.\ we do not enlarge the $v_{\rm min}$ interval using the energy resolution. Having determined $\overline{\tilde\eta_{0i}} = \hat{R}_{0i}/\mathcal{A}^{SI}_{{\rm O},i}$ for $i=4,5,6$ using O only in $\mathcal{A}^{SI}_{{\rm O},i}$, we estimate the Ca contribution to bins $j=1,2,3$ as $R^{SI}_{{\rm Ca},j} = \mathcal{A}^{SI}_{{\rm Ca},j} \overline{\tilde\eta_{0,j+3}} $, where $\mathcal{A}^{SI}_{{\rm Ca},j}$ contains only Ca. Then to reduce the effect of the propagation of errors in subtracting the Ca contribution, we combine the three low-energy bins into one, obtaining for it $\overline{\tilde\eta_{0}}=\sum_{j=1}^{3}(\hat{R}_{0j}-R^{SI}_{{\rm Ca},j})/ \sum_{j=1}^{3}\mathcal{A}^{SI}_{{\rm O},j}$.
 \begin{figure}[t]
\includegraphics[width=8.0cm]{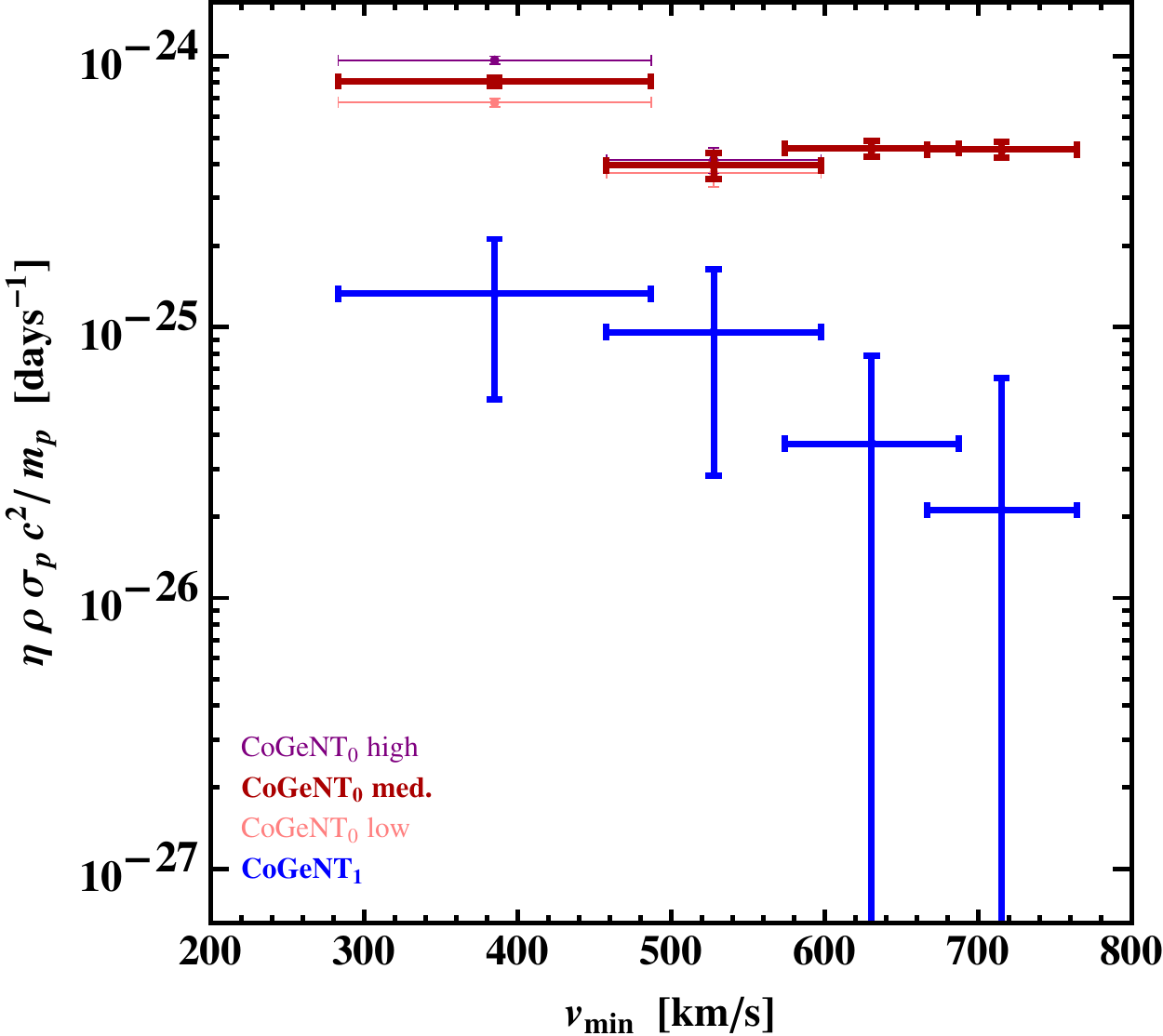}
\caption{CoGeNT measurement of the unmodulated part of the velocity integral $\eta(v_{\rm min})$ plus background,  $\eta_0 +b_0$, and of the
modulated part,  $\eta_1$,  of the velocity integral $\eta(v_{\rm min})$ as a function of  $v_{\rm min}$, for a WIMP with spin-independent isospin-symmetric couplings and mass of 9 GeV.}
\label{fig:1}
\end{figure}

\section{IV. Our results}

The figures show our results for a WIMP with spin-independent couplings and mass 9 GeV. To compare with the corresponding figures in~\cite{Frandsen:2011gi}, which have the vertical axis in units of inverse days, we multiply $\tilde{\eta}$ by the square of the speed of light $c^2$. We warn the reader that the quantity plotted may seem to be the number of WIMPs impinging on the detector per day, but it is actually not. Notice that $\eta$ in the label of the vertical axis stands for either $\eta_0$ or $\eta_1$ depending on the experiment. We plot both the modulated and unmodulated parts of $\eta$ in the same figure to be able to compare them. In all realistic cases we should have $\eta_1$ sufficiently smaller than  $\eta_0$.

Fig.~\ref{fig:1} shows the CoGeNT measurement of the unmodulated part of the velocity integral $\eta(v_{\rm min})$ plus background,  $\eta_0 +b_0$ (high, medium and low), and of the
modulated part,  $\eta_1$,  of the velocity integral $\eta(v_{\rm min})$ as a function of  $v_{\rm min}$, for a WIMP  isospin-symmetric couplings.  It clearly shows that the  modulation amplitude of the CoGeNT data is large with respect to the average, certainly larger than the few percent modulation in usual halo models. If the unmodulated CoGeNT rate at 
 \begin{figure}[t]
\includegraphics[width=8.0cm]{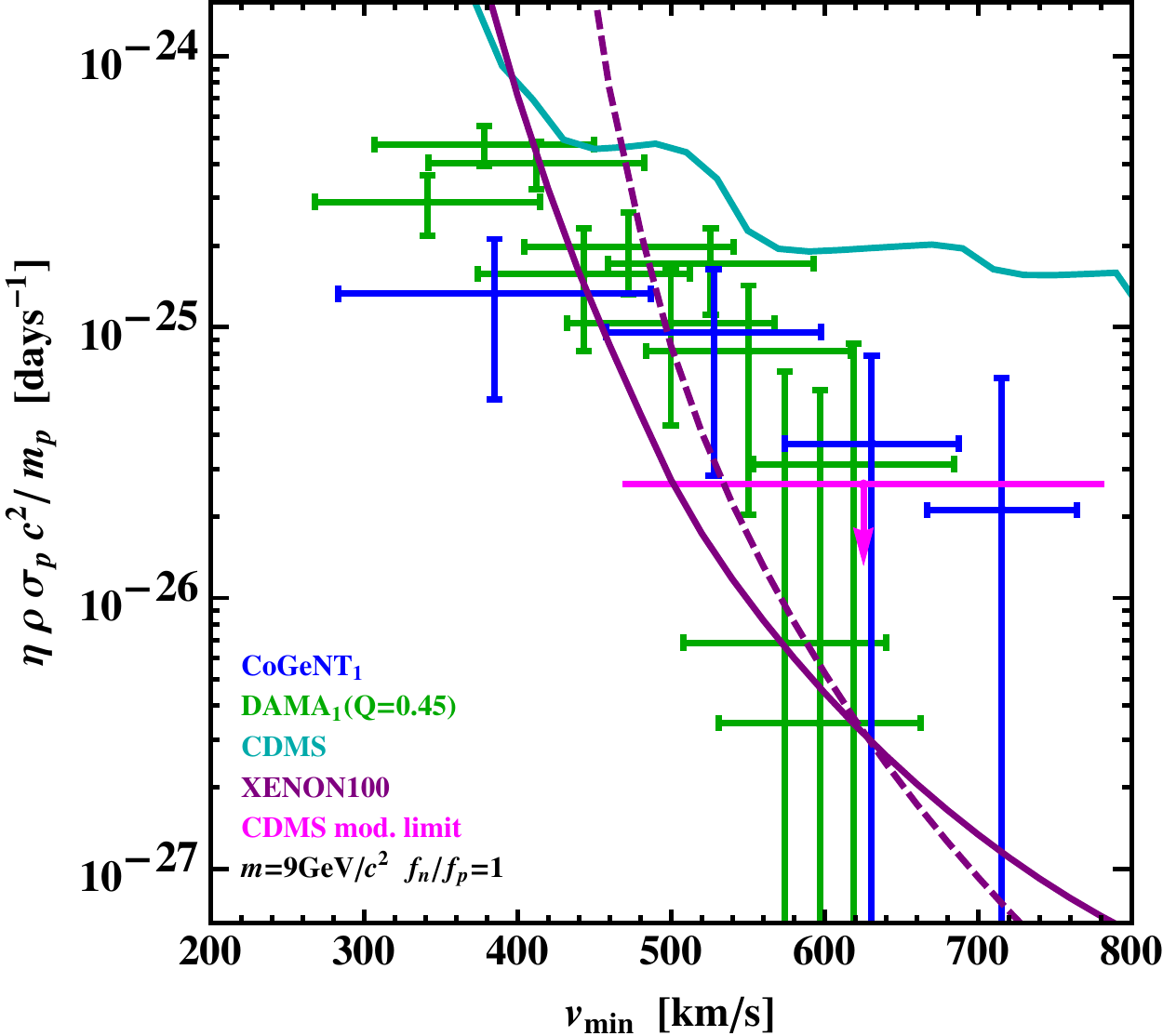}
\caption{CoGeNT and DAMA measurements  of and CDMS bounds on the modulated part $\eta_1$,  and CDMS and XENON 100 (solid line for the latest data) bounds on the unmodulated part $\eta_0$ of  $\eta(v_{\rm min})$,  as a function of  $v_{\rm min}$.  
Here $Q_{\rm Na} =$0.45, $m=$ 9 GeV. For this case of spin-independent isospin-symmetric couplings, the XENON100 and CDMS modulation bounds exclude all but the lowest energy CoGeNT and DAMA bins.}
\label{fig:2}
\end{figure}
 \begin{figure}[t]
\includegraphics[width=8.0cm]{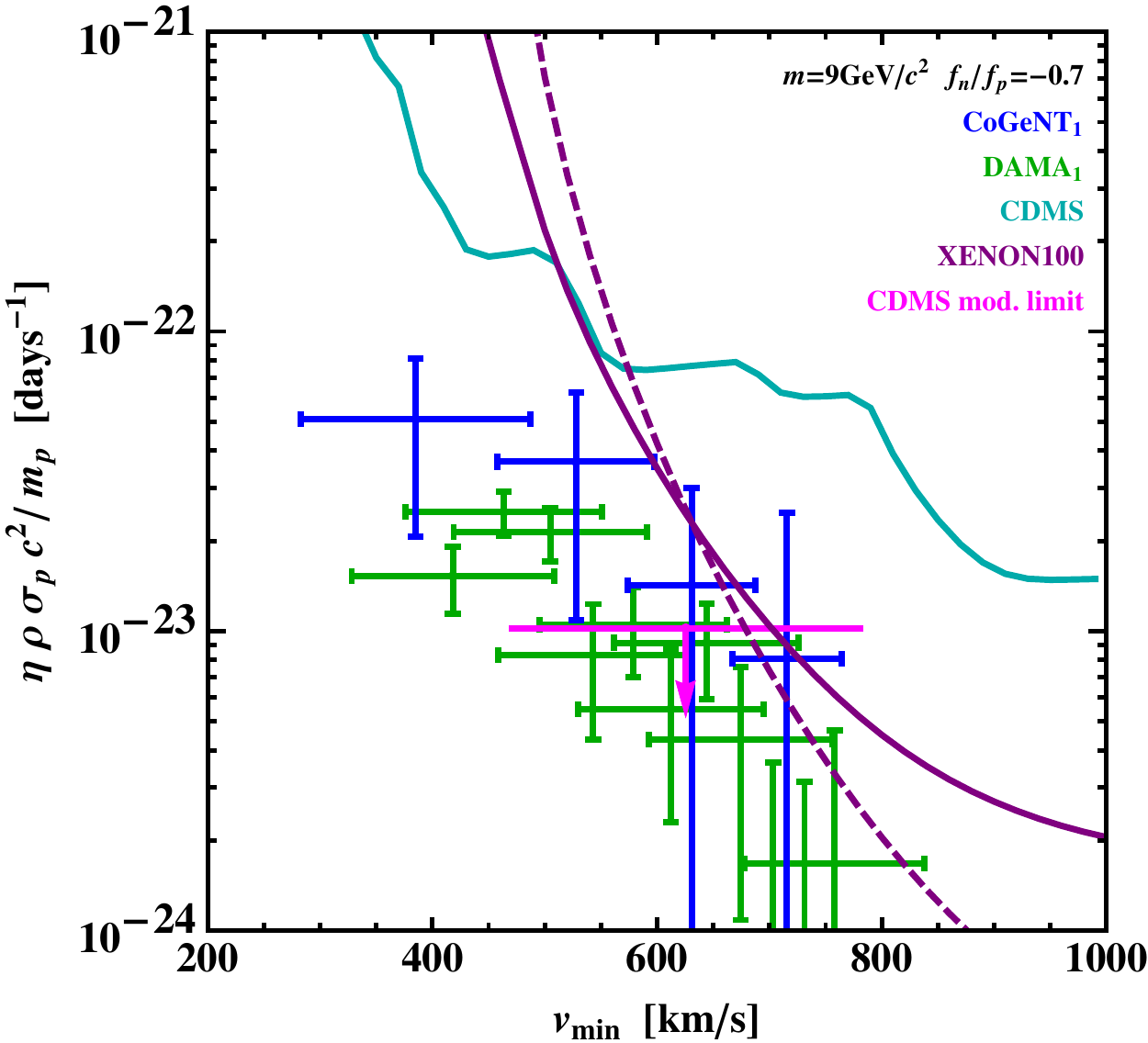}
\caption{As in Fig.~\protect\ref{fig:2} but for $Q_{\rm Na} =$0.30 and $f_n/f_p \simeq -0.7$.
The first two CoGeNT 
and the first six DAMA energy bins are compatible with XENON100 bounds but the CDMS modulation constraint  exclude all but the lowest points (since both CoGeNT and CDMS use Ge, points and limit move together).
}
\label{fig:3}
\end{figure}
\begin{figure}[t]
\includegraphics[width=8.0cm]{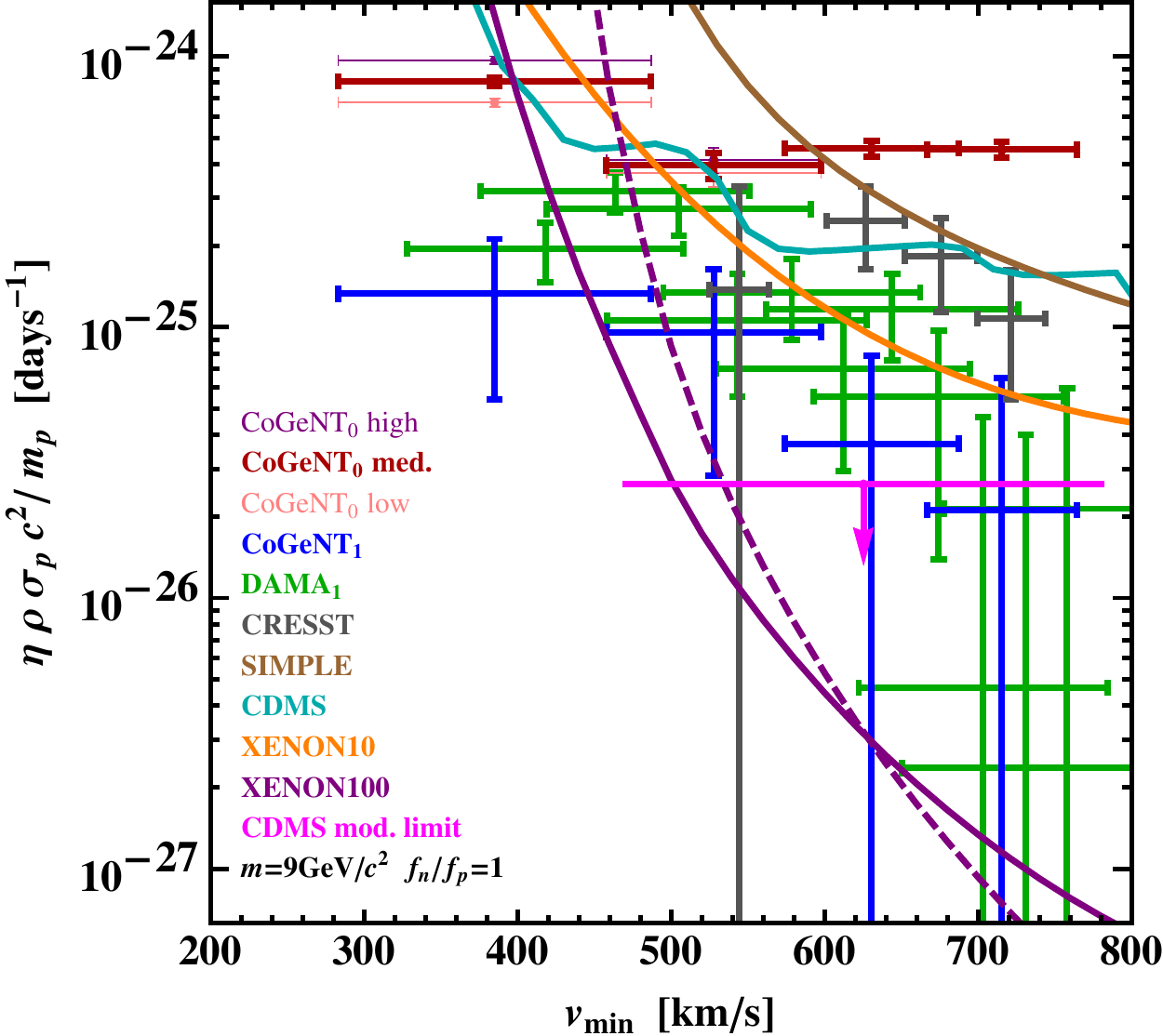}
\caption{Measurements and upper bounds on the unmodulated and modulated part of the velocity integral $\eta(v_{\rm min})$ as a function of  $v_{\rm min}$, for $m=$ 9 GeV.  For this case of spin-independent isospin-symmetric couplings,
the XENON100 and CDMS modulation bounds exclude all DAMA and all but the lowest energy CoGeNT bins.}
\label{fig:4}
\end{figure}
\begin{figure}[t]
\includegraphics[width=8.0cm]{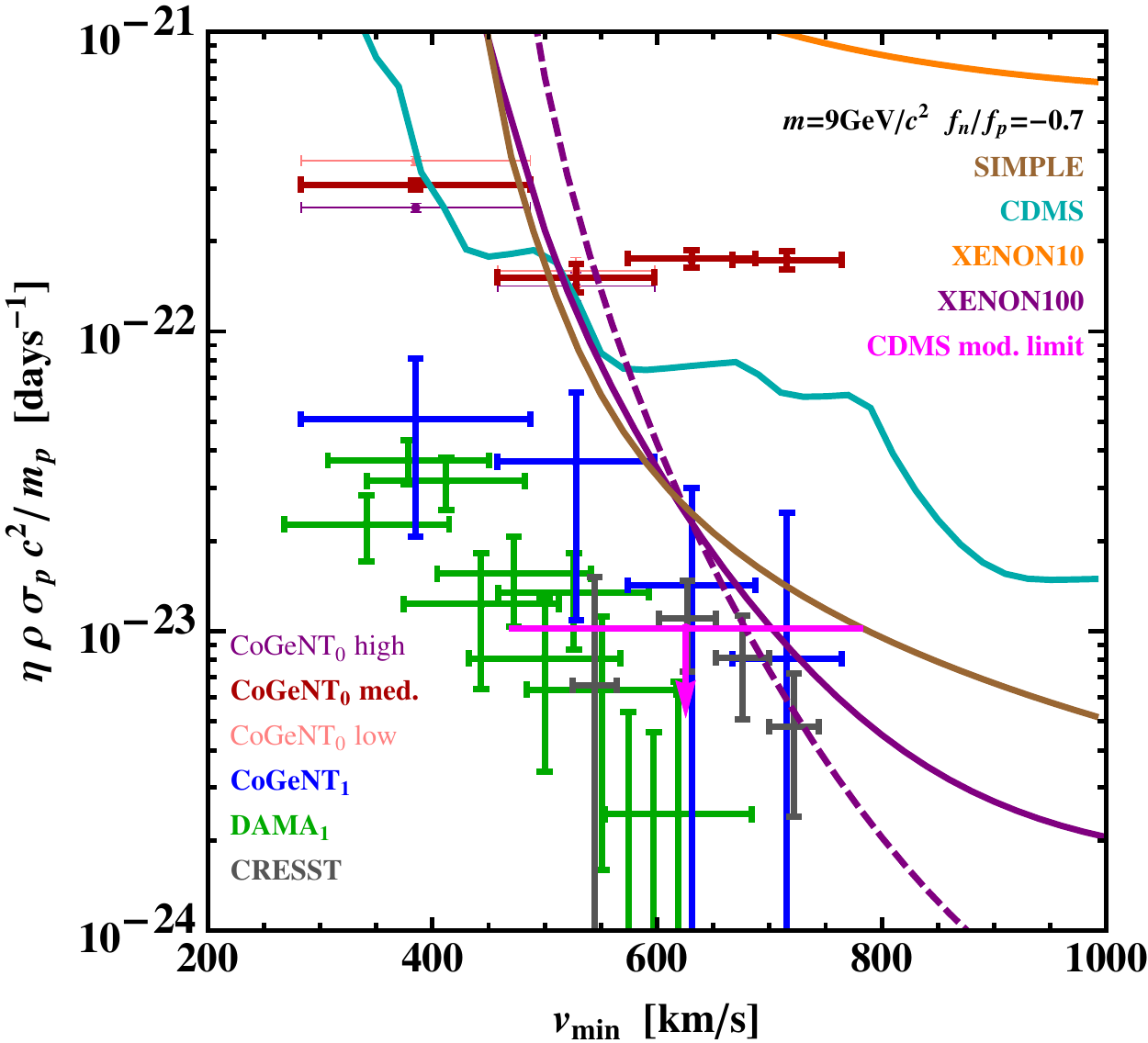}
\vspace{-0.2cm}
\caption{Same as Fig.~\protect\ref{fig:4} but for isospin-violating couplings $f_n/f_p=-0.7$ and DAMA quenching factor $Q_{\rm Na}=0.45$. In this case all DAMA and the lowest $v_{\rm min}$ CoGeNT points are allowed by all the bounds we consider except the CDMS modulation constraint, which excludes all but the lowest bins.
}
\label{fig:5}
\end{figure}
high recoil energies is subtracted throughout the energy range, the relative modulation amplitude would have to be even higher, $\sim30$\%. 

In Figs.~\ref{fig:2} and \ref{fig:3} we compare  just the CoGeNT and DAMA measurements  of, and CDMS upper bounds on, the modulated part,  $\eta_1$, of the velocity integral $\eta(v_{\rm min})$, 
as well as the CDMS and XENON100 (the continuous line corresponds to the latest data) bounds on the unmodulated part $\eta_0$ of   $\eta(v_{\rm min})$, as a function of  $v_{\rm min}$. 
Figs.~\ref{fig:4} and~\ref{fig:5} show all the measurements and upper bounds on $\tilde\eta_0$ and $\tilde\eta_1$ included in this paper.  

Figs.~\ref{fig:2} to \ref{fig:5} show that the DAMA and CoGeNT measurements of the modulation amplitude $\eta_1$ are compatible with each other. 

 In Fig.~\ref{fig:2} and Fig.~\ref{fig:5} 
we used $Q_{\rm Na} =$0.45, considered a high value relative to measurements, instead of the usual 0.3 used 
in the other figures. The effect of changing $Q_{\rm Na}$ from 0.3 to 0.45 can be seen by comparing the relative position of the DAMA points (green) with respect to the CoGeNT points (blue) when going from Fig.~\ref{fig:4} to Fig.~\ref{fig:2} (both with $f_n/f_p =1$) and from Fig.~\ref{fig:3} to Fig.~\ref{fig:5} (both with $f_n/f_p =-0.7$).
Since a particular detected energy is $Q E$, a change in $Q$ from $Q_{\rm old}$ to $Q_{\rm new}$ corresponds
to a horizontal shift of the data points in $E$ to $E_{\rm new}= (Q_{\rm old}/  Q_{\rm new}) E_{\rm old}$ and a vertical shift of the data points by a factor $Q_{\rm new}/Q_{\rm old}$. Thus, an increase in $Q$ leads to a diagonal leftwards and upwards shift of the points to lower  $v_{\rm min}$ by a factor $\sqrt{Q_{\rm old}/  Q_{\rm new}}$ and a higher $\rho \sigma_p \eta$  by  the factor.
($Q_{\rm old}/  Q_{\rm new}$) (a decrease in $Q$ causes a diagonal shift to the right and down in the diagram of the affected points).  
  
 For the case  of isospin-symmetric couplings of Fig.~\ref{fig:2}  and~\ref{fig:4}  the XENON100 and CDMS modulation bounds exclude all but the lowest energy CoGeNT and DAMA bins. Although a larger value of $Q_{\rm Na}$ in  Fig.~\ref{fig:2} shift the two lowest DAMA points outside the XENON 100 bounds,  the tension between CoGeNT and DAMA on one side and XENON100 and CDMS on the other is strong. Varying the WIMP mass from 6 to 12 GeV does not improve the situation.

The tension is alleviated for isospin-violating couplings $f_n/f_p=-0.7$, shown in Figs.~\ref{fig:3} and~\ref{fig:5}, especially if the DAMA Na quenching factor is taken as $Q_{\rm Na}=0.45$ (in Fig.~\ref{fig:5}). 
In this case, the fist  (lowest energy) two  CoGeNT data points  and the first either six (in Fig.~\ref{fig:3}) or all (in Fig.~\ref{fig:5}) DAMA points are compatible with all the bounds we consider,  except the CDMS modulation limit.  Since CDMS and CoGeNT both use Ge, the tension of the higher energy CoGeNT bins with the CDMS modulation constraint remains. Even with $Q_{\rm Na}=0.45$ (see Figs.~\ref{fig:2} and~\ref{fig:5}) the CDMS modulation bound rejects all but the two lowest energy DAMA points.
It is therefore of the utmost interest that CDMS extend their modulation analysis to lower energies, so as to confirm or exclude the spin-independent interpretation of the CoGeNT and DAMA annual modulations over the full $v_{\rm min}$ range.

  Notice that the  $f_n/f_p=-0.7$  choice diminishes not only the WIMP-Xenon coupling, but also the WIMP  couplings with Ge and Na. Comparing e.g.  Figs.~\ref{fig:2} and~\ref{fig:3} we see that the coupling $[Z+ (A-Z)(f_n/f_p)]^2$ in  Eq.~\ref{sigma} of the WIMP with Ge diminished, thus $\tilde\eta$ increased, by a factor of about 500. For Na the factor is smaller (compare  e.g. Figs.~\ref{fig:3} and~\ref{fig:4}), about 70.  Thus, with $f_n/f_p=-0.7$ the DAMA modulation points are below the CoGeNT data points (while they are above with $f_n/f_p=1$).

Fig.~\ref{fig:4}  and~\ref{fig:5} show that the CRESST II measurements of the unmodulated part, $\eta_0$
of  the velocity integral $\eta(v_{\rm min})$ (in dark grey) are superposed to the DAMA and CoGeNT measurements of the modulated part $\eta_1$, while in realistic models it should be that  $\eta_1 < \eta_0$. Thus we find that CRESST II results are incompatible with the CoGeNT and DAMA modulation data.

In conclusion,  in this paper we extended the halo-independent method of Fox, Liu, and Weiner to include energy resolution and efficiency with arbitrary energy dependence, making it more suitable for experiments to use in presenting their results. We also show data comparisons for spin independent 
WIMPs using this method.

\section*{Acknowledgments}

We thank Chris Savage for extensive discussions at the beginning of this work, and Peter Sorensen for providing us with the list of event energies in XENON10~\cite{Angle:2011th}.
P.G. was supported in part by NSF grant PHY-1068111, and thanks KIAS for hospitality during part of his sabbatical year. G.G. was supported in part by DOE grant DE-FG03-91ER40662, Task C and thanks the Aspen Center for Physics and the NSF Grant 1066293 for hospitality during the editing of this paper.

\end{document}